\documentclass[prb,twocolumn,showpacs]{revtex4}

\usepackage{graphicx}
\usepackage{amsmath}
\usepackage{amsfonts}
\usepackage{amssymb}
\usepackage{dcolumn}
\usepackage{bm}

\begin{document}

\title{Mott-Hubbard versus charge-transfer behavior in LaSrMnO$_4$\\studied via optical conductivity}

\author{A. G\"{o}ssling, M.~W.~Haverkort, M. Benomar, Hua Wu, D. Senff, T. M\"{o}ller, M. Braden, J.~A.~Mydosh, and M. Gr\"{u}ninger}
\affiliation{$II.$ Physikalisches Institut, Universit\"{a}t zu K\"{o}ln, Z\"{u}lpicher Str.\ 77, 50937 K\"{o}ln, Germany}

\begin{abstract}
Using spectroscopic ellipsometry, we study the optical conductivity $\sigma(\omega)$ of insulating LaSrMnO$_4$ in the
energy range of 0.75-5.8 eV from 15 to 330 K.\@ The layered structure gives rise to a pronounced anisotropy.
A multipeak structure is observed in $\sigma_1^a(\omega)$ ($\sim2$, 3.5, 4.5, 4.9, and 5.5 eV), while only one peak
is present at 5.6 eV in $\sigma_1^c(\omega)$.
We employ a local multiplet calculation and obtain (i) an excellent description of the optical data,
(ii) a detailed peak assignment in terms of the multiplet splitting of Mott-Hubbard and charge-transfer
absorption bands, and (iii) {\em effective} parameters of the electronic structure, e.g., the on-site Coulomb
repulsion $U_{\rm eff}$=2.2 eV,
the in-plane charge-transfer energy $\Delta_a$=4.5 eV, and the crystal-field parameters for the $d^4$ configuration
($10\,Dq$=1.2 eV, $\Delta_{eg}$=1.4 eV, and $\Delta_{t2g}$=0.2 eV).
The spectral weight of the lowest absorption feature (at 1-2 eV) changes by a factor of 2 as a function of temperature,
which can be attributed to the change of the nearest-neighbor spin-spin correlation function across the Neel temperature
$T_N$=133 K.\@
Interpreting LaSrMnO$_4$ effectively as a Mott-Hubbard insulator naturally explains this strong temperature dependence,
the relative weight of the different absorption peaks, and the pronounced anisotropy.
By means of transmittance measurements, we determine the onset of the optical gap $\Delta^a_{\rm opt}$ = 0.4-0.45 eV
at 15 K and 0.1-0.2 eV at 300 K.\@ Our data show that the crystal-field splitting is too large to explain
the anomalous temperature dependence of the $c$-axis lattice parameter by thermal occupation of excited
crystal-field levels. Alternatively, we propose that a thermal population of the upper Hubbard band gives rise to
the shrinkage of the $c$-axis lattice parameter.
\end{abstract}


\date{September 21, 2007}
\maketitle

\section{Introduction}

The insulating behavior of many transition-metal (TM) compounds with a partially filled $3d$ shell is a clear manifestation
of strong electronic correlations. In the Zaanen-Sawatzky-Allen scheme,\cite{ZSA} these correlated insulators
are categorized into Mott-Hubbard (MH) and charge-transfer (CT) types. In the former, the on-site Coulomb repulsion $U$ is
larger than the band width $W$; thus, the conduction band splits into lower and upper Hubbard bands (LHB and UHB,
see Fig.\ 1). At half filling, the lowest electronic excitation is from LHB to UHB, $d^n d^n \rightarrow d^{n-1} d^{n+1}$,
where $d^n$ denotes a TM ion with $n$ electrons in the $3d$ shell.
In a CT insulator, $U$ is larger than the charge-transfer energy $\Delta$; thus, the LHB is pushed below the highest occupied
band of the ligands, e.g., O $2p$ (see Fig.\ 1). Here, the lowest electronic excitation is from O $2p$ to UHB,
$p^6 d^n  \rightarrow p^5 d^{n+1}$. Typically, early TM compounds are of MH type, whereas late ones such as the
high-$T_c$ cuprates belong to the CT class.\cite{arima93a}

\begin{figure}[b!]
     \includegraphics[width=0.7\columnwidth,clip]{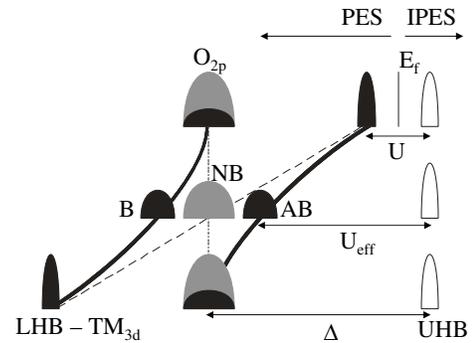}
    \caption{Sketch of the density of states [photoemission (PES) and inverse photoemission (IPES) spectra] for a
    single, half-filled $3d$ orbital and degenerate, full O $2p$ orbitals.
    The on-site Coulomb repulsion $U$ increases from top to bottom, whereas the charge-transfer energy $\Delta$ is
    assumed to be constant. $E_F$ denotes the Fermi level.
    Top: Mott-Hubbard insulator for $U\! \ll \! \Delta$.
    Bottom: charge-transfer insulator for $U\! \gg \! \Delta$.
    In the absence of hybridization, the bands follow the straight lines with increasing $U$.
    With hybridization, one has to distinguish bonding (B), antibonding (AB, both black), and nonbonding bands (NB, gray).
    }
    \label{fig1}
\end{figure}

For manganites with Mn$^{3+}$ $3d^4$ configuration, e.g., LaMnO$_3$ or LaSrMnO$_4$, the characterization is not as straightforward.
The analysis of photoemission data indicates $U \gtrsim \Delta$, i.e., a CT character.\cite{mizokawa96}
Yet, most theoretical approaches employ an {\rm effective} Hubbard model (see below).
Both MH and CT types have been proposed on the basis of optical data.\cite{arima93a,moritomo95a,jung97a,quijada,lee07a,tobe01a,kovaleva04a,ahn00}
For $U\gg \Delta$ or $U\ll \Delta$,
there are different ways to disentangle MH and CT excitations in the optical conductivity.
First, the spectral weight of CT excitations is larger, e.g., $\sigma_1(\omega) \sim$500 $(\Omega$cm$)^{-1}$ for MH and
$\sim$5000 $(\Omega$cm$)^{-1}$ for CT excitations in LaTiO$_3$.\cite{arima93a} This can be attributed to the TM$_{3d}$-O$_{2p}$
hopping amplitude $t_{pd}$: $\sigma_1(\omega)\propto t_{pd}^2$ for CT and $\propto t_{pd}^4/\Delta^2$ for MH excitations (see below).
Second, the polarization dependence may be very different, depending on the crystal structure.
In layered LaSrMnO$_4$, the interlayer Mn-Mn hopping is strongly suppressed; thus, the contribution
of MH excitations should be negligible in $\sigma_1^c(\omega)$. In contrast, CT excitations contribute
to both $\sigma_1^a(\omega)$ and $\sigma_1^c(\omega)$ due to the octahedral O coordination of the Mn sites.
Third, the spin and orbital selection rules are different, giving rise to a different behavior of the spectral
weight as a function of temperature $T$.\cite{ahn00,tobe01a,kovaleva04a,lee05,lee02a,kim04a,rauer,miyasaka02,tsvetkov04a,khaliullin04a,oles05a}
As an example, we consider a MH insulator with a single $3d$
band and one electron per TM site. In the case of ferromagnetic order, the spectral weight of MH excitations is
zero due to the Pauli principle (neglecting spin-orbit coupling, the total spin is conserved in an optical excitation).
Thus, one expects a drastic change of the spectral weight upon heating across the magnetic ordering temperature.
In contrast, the O $2p$ band is completely filled and the spectral weight of CT excitations is independent
of the magnetic properties.
For comparison with experimental data, the orbital multiplicity has to be taken
into account and the optical spectra of MH excitations contain valuable information on orbital occupation
and nearest-neighbor spin and orbital correlations.
This kind of analysis has been applied to a number of different TM compounds
(Mn, V, Ru, Mo).\cite{ahn00,tobe01a,kovaleva04a,lee05,lee02a,kim04a,rauer,miyasaka02,tsvetkov04a,khaliullin04a,oles05a}
In LaMnO$_3$, there is some disagreement on the experimental side:
LaMnO$_3$ has been interpreted as MH type due to a pronounced $T$ dependence observed by spectroscopic ellipsometry\cite{kovaleva04a}
and as CT type due to the absence of a significant $T$ dependence across the Neel temperature $T_N$ in reflectivity
measurements.\cite{tobe01a}
Note that it is still unclear in which compounds this kind of analysis can be applied. For instance, in YTiO$_3$, the lowest MH excitation
shows a strong $T$ dependence but only a weak dependence on the magnetic properties.\cite{goessling06}
However, our analysis of LaSrMnO$_4$ fully corroborates the intimate relation between spectral weight and magnetism reported
by Kovaleva {\em et al.}\cite{kovaleva04a} for LaMnO$_3$.

Does this prove that LaMnO$_3$ and LaSrMnO$_4$ are of MH type?
The $d^5$ configuration is particularly stable. Thus, the $d^4$ configuration has a strong $d^5 \underline{L}$
contribution, where $\underline{L}$ denotes a ligand hole and the O $2p$ orbitals certainly play an important role.
As stated above, with $U\gtrsim \Delta$,\cite{mizokawa96} the manganites may be categorized as CT type.
However, for the analysis of, e.g., optical data, it is important to consider the symmetry.
Let us discuss the simplified case of a single, half-filled $3d$ band and degenerate, full O $2p$ bands. Due to hybridization,
one has to distinguish bonding (B), antibonding (AB), and nonbonding (NB) bands below the Fermi level (see Fig.\ 1).
The AB band is the highest occupied one. It has mainly TM character for $U\! \ll \! \Delta$ and mainly O character
for $U\! \gg \! \Delta$. For the intermediate situation $U\sim \Delta$, the character is strongly mixed.
However, in the manganites with a less than half-filled $3d$ shell, the {\em symmetry} of the AB band is determined by
the symmetry of the $3d$ band, i.e., it can be described in terms of a Wannier orbital with $d$ symmetry centered
around a Mn site. In particular, the spin and orbital selection rules mentioned above for a MH insulator apply also
to the AB band. These selection rules are most important for understanding the optical data. For instance, in LaSrMnO$_4$,
one expects that the AB band shows the pronounced anisotropy and the strong temperature dependence discussed above
for the MH case.
Thus, {\em ``MH or CT''} is not the essential question for $U\sim \Delta$ and strong hybridization.
In this sense, we analyze our data in terms of an {\em effective} Hubbard model with a rather small value
of $U_{\rm eff}$=2.2 eV.\@
Excitations from NB to UHB are treated as CT and from AB to UHB as MH.\@ We find that these effective
MH excitations are significantly lower than the CT excitations.
The effective model does not contain the B band, which is important for optical or photoemission measurements
at higher energies.
However, we fully take into account the multiplet splitting of the MH and CT bands
caused by the multiorbital structure of the $3d$ shell.

Here, we focus on LaSrMnO$_4$, which crystallizes in the layered K$_2$NiF$_4$ structure with tetragonal symmetry
$I4/mmm$.\cite{reutler03a} The lattice constants at room temperature are $a$=3.786 \AA$\;$ and $c=$13.163 \AA.
All Mn ions have a nominal valence of $+3$ ($3d^4$ with spin $S$=2). Antiferromagnetic order has been observed below
$T_N$=133 K.\cite{larochelle05a,senff05a,krienerthesis}
Neglecting hybridization, the four $3d$ electrons occupy the $xy$, $yz$, $zx$, and $3z^2$-$r^2$ orbitals with parallel spins,
while $x^2$-$y^2$ is empty.\cite{wu04a}
Similar to LaMnO$_3$, also the character of LaSrMnO$_4$ -- MH vs CT -- has been discussed
controversially.\cite{moritomo95a,park01a,kuepper06a,kuepper06b,merz06a,merz06b,lee07a}
Previous optical studies reported $\sigma_1^a(\omega)$ for 300 K (Ref.~\onlinecite{moritomo95a}) and very recently for 10 K.\cite{lee07a}
By means of spectroscopic ellipsometry, we determine the optical conductivity tensor $\sigma(\omega)$ between 0.75 and 5.8 eV
as a function of temperature. Using a local multiplet calculation, we interpret the observed absorption bands in terms of
effective Mott-Hubbard and charge-transfer excitations. This assignment utilizes the points raised above, i.e., the pronounced anisotropy,
the difference in spectral weight, and the strong temperature dependence of the spectral weight.
Using transmittance measurements, we determine the optical gap $\Delta_a$=0.4-0.45 eV at 15 K and 0.1-0.2 eV at 300 K.\@
Our data indicate that the anomalous shrinkage of the $c$-axis lattice parameter with increasing temperature cannot be
attributed to a thermal occupation of {\em local} crystal-field levels\cite{senff05a,daghofer06a,daghofer06b}
but rather to a thermal population of the UHB.\@

Section II describes the experimental details. Our optical data and a detailed analysis of the spectral weight are reported
in  Sec.\ III.\@ The local multiplet calculation, the effective parameters of the electronic structure, and the
calculation of $\sigma(\omega)$ based on the multiplet calculation are discussed in Sec.\ IV.\@
Finally, in Sec.\ V, we discuss the peak assignment in terms of the multiplet splitting of MH and CT absorption bands,
the relationship between the spectral weight and superexchange, and the anomalous thermal expansion.

\section{Experimental}

\begin{figure}[t!]
     \includegraphics[width=0.95\columnwidth,clip]{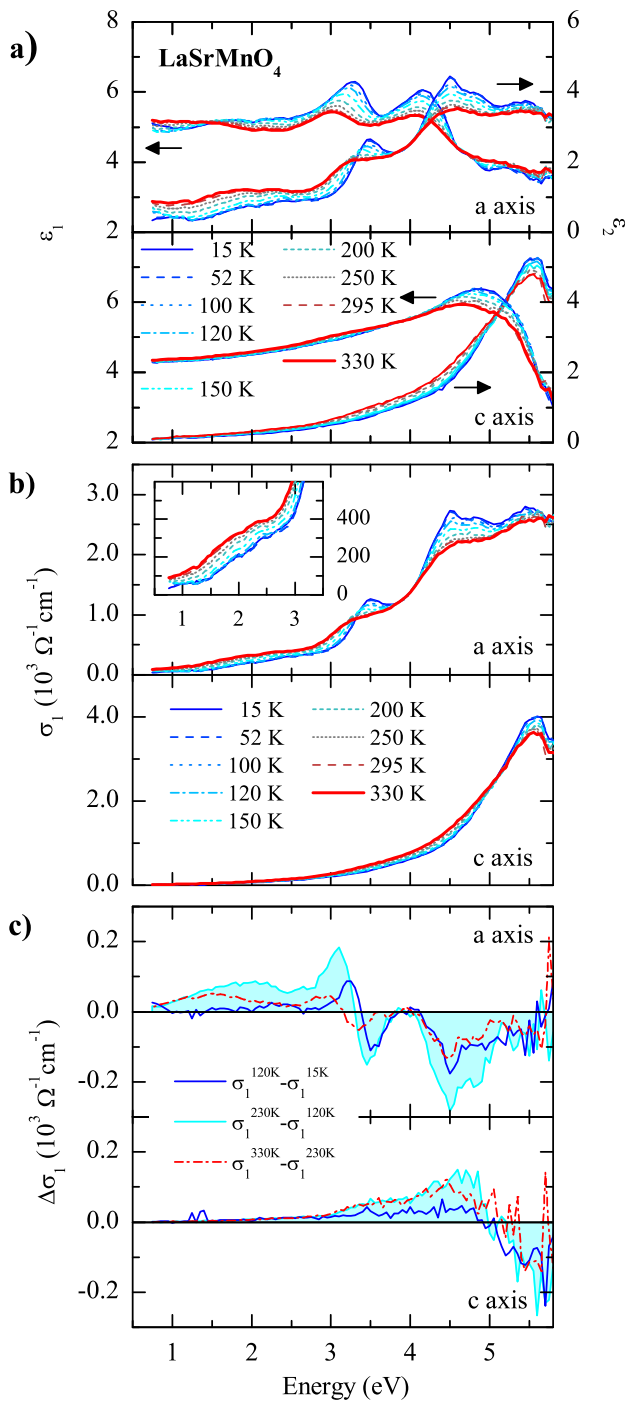}
    \caption{(Color online) [(a) and (b)] Dielectric constant and optical conductivity
    of LaSrMnO$_4$ for the $a$ and $c$ directions between 0.75 and 5.80 eV for different
    temperatures. (c) Change of the optical conductivity: $\sigma_1(330\;\mathrm{K})$-$\sigma_1(230\;\mathrm{K})$,
    $\sigma_1(230\;\mathrm{K})$-$\sigma_1(120\;\mathrm{K})$, and $\sigma_1(120\;\mathrm{K})$-$\sigma_1(15\;\mathrm{K})$.
    }
    \label{fig2}
\end{figure}

Single crystals of LaSrMnO$_4$ have been grown using the floating zone technique
following Ref.\ \onlinecite{reutler03a}.
The sample quality and stoichiometry were checked using polarization microscopy,
neutron diffraction, and x-ray diffraction.
The two nonvanishing, complex entries $\varepsilon^a$ and $\varepsilon^c$ of the dielectric
tensor for tetragonal symmetry were determined using a rotating-analyzer ellipsometer
(Woollam VASE) equipped with a retarder between polarizer and sample.
By measuring on a polished $ac$ surface, we determined the normalized M\"{u}ller matrix elements
$m_{ij}^k=M_{ij}^k/M_{11}^k$ ($i$=1-3, $j$=1-4, $k$=1-2),\cite{azzam87a} where $k$=1 (2) corresponds
to measurements with the $a$ ($c$) axis perpendicular to the plane of incidence.
The angle of incidence was 70$^\circ$.
We obtained $\varepsilon^a$ and $\varepsilon^c$ by fitting the nonvanishing elements
$m_{12}^k$, $m_{21}^k$, $m_{33}^k$, and $m_{34}^k$.\cite{goesslingthesis}
We have checked that the results fulfill the Kramers-Kronig consistency.
The ellipsometric measurements have been performed from 15 to 330 K
in a UHV cryostat with $p<10^{-9}$ mbar. The effect of the cryostat windows has
been determined using a standard Si wafer.
Our data are consistent with spectra of Refs.\ \onlinecite{moritomo95a} and \onlinecite{lee07a}.

In order to determine the optical gap, we measured the transmittance from 5 to 300 K
using a Fourier-transform spectrometer (Bruker IFS 66v). The sample was
approximately 70 $\mu$m thick and has been prepared in the same way as described above.

\begin{figure}[tb]
     \includegraphics[width=0.95\columnwidth,clip]{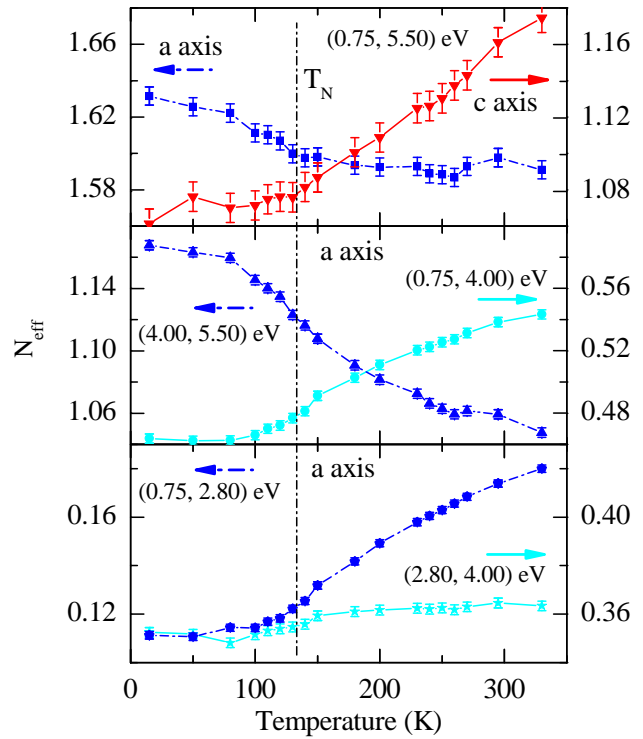}
    \caption{(Color online) Top: effective carrier concentration $N_{\rm eff}(\omega_1, \omega_2)$
    for the $a$ and $c$ directions with $\omega_1$=0.75 eV and $\omega_2$=5.5 eV.\@
    Middle and bottom: $N_{\rm eff}$ for the $a$ direction for different energy ranges.
    }
    \label{fig3}
\end{figure}

\section{Results}

\begin{figure}[tb]
     \includegraphics[width=0.90\columnwidth,clip]{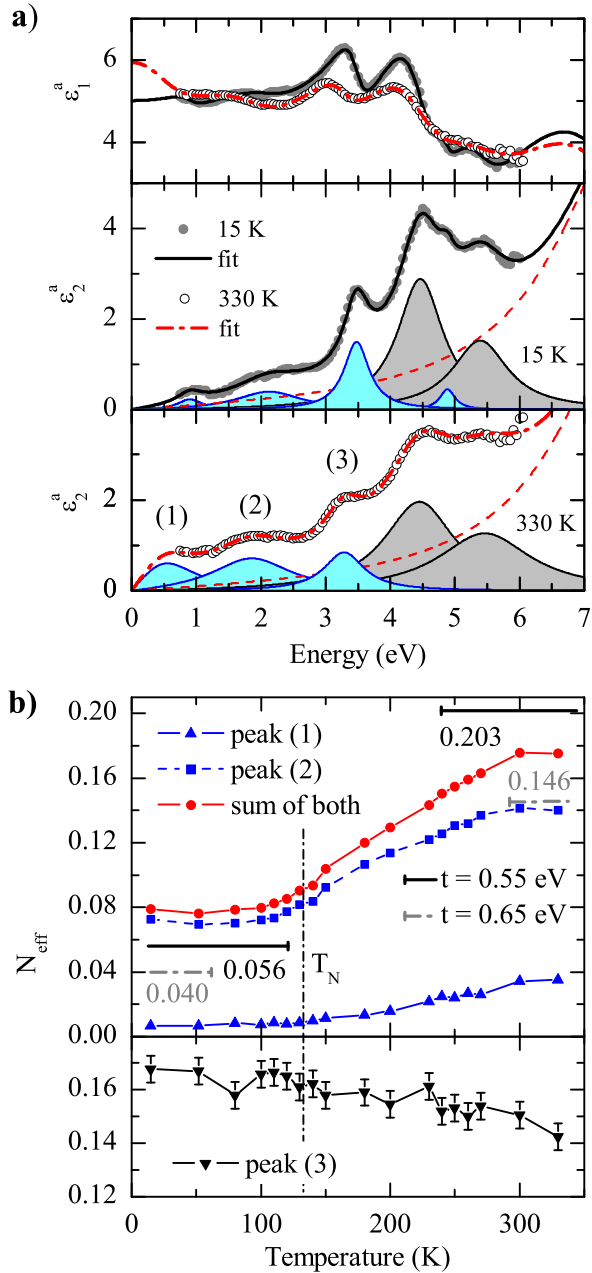}
    \caption{(Color online) (a) Lorentzian fit of $\epsilon_1^a$ (top panel) and
    $\epsilon_2^a$ (middle and bottom) for $T$=15 and 330 K.\@
    (b) $N_{\rm eff}$ of peaks (1)-(3) as obtained from the Drude-Lorentz fit.
    The horizontal lines indicate theoretical estimates of $N_{\rm eff}$ for $T \ll T_N$ and
    $T \gg T_N$ as derived from the kinetic energy for an effective Mn-Mn hopping amplitude
    $t$=0.55-0.65 eV (see Sec.\ V C).
    }
    \label{fig4}
\end{figure}

\subsection{Ellipsometry and interband excitations}

In Fig.\ \ref{fig2}(a), we plot $\varepsilon^{l}=\varepsilon^{l}_1 + \imath \varepsilon^{l}_2$ ($l$=$a$,$c$) from 0.75 to 5.8 eV.\
For convenience, the real part $\sigma_1=(\omega/4 \pi)\,\varepsilon_2$ of the optical conductivity is displayed
in Fig.\ \ref{fig2}(b). We find a striking anisotropy. In particular, there is only one strong peak at 5.6 eV in
$\sigma_1^c$, while a multipeak structure is present in $\sigma_1^a$ (peaks at $\sim$2, 3.5, 4.5, 4.9, and
5.5 eV). All peaks show a strong temperature dependence. With increasing $T$, $\sigma_1^a$ increases below
$\sim$3 eV and decreases above $\sim$4 eV.\@ In particular, spectral weight (SW) is transferred from $3.5$ to $3.0$ eV
and from $4.5$ to $2.0$ eV [see Fig.\ \ref{fig2}(c)].

We analyze the $T$ dependence of the SW using the effective carrier concentration $N_{\rm eff}$,
\begin{eqnarray} \label{eq:Neff}
N_{\rm eff}(\omega_{1},\omega_{2})=\frac{2mV}{\pi e^2}\int_{\omega_{1}}^{\omega_{2}} \sigma_1(\omega) d\omega
\end{eqnarray}
where $m$ denotes the electron mass, $V$ the unit cell volume, and $e$ the electron charge.
Equation (\ref{eq:Neff}) translates into the $f$-sum rule for $\omega_{1}\! \rightarrow \! 0$ and
$\omega_{2}\! \rightarrow \! \infty$.\cite{dressel02a}
As shown in Fig.\ \ref{fig3}, the total spectral weight $N_{\rm eff}(0.75\,\mathrm{eV},5.5\,\mathrm{eV})$ decreases (increases)
with increasing $T$ for the $a$ ($c$) direction. In $\sigma_1^a$, we find an isosbestic point at
$\omega_i \! \approx \! 4.0$ eV.\@ The corresponding transfer of SW is evident from the comparison of
$N_{\rm eff}(0.75\,\mathrm{eV},4.0\,\mathrm{eV})$ and $N_{\rm eff}(4.0\,\mathrm{eV},5.5\,\mathrm{eV})$ (see middle
panel of Fig.\ \ref{fig3}). The transfer of SW sets in roughly 30 K below $T_N$=133 K, but
the curves posses an inflection point approximately at $T_N$.

The direct integration of $\sigma_1(\omega)$ in Eq.\ 1 has the advantage to be model independent. For a more
detailed analysis of the $T$ dependence of the individual absorption bands, we fit
$\varepsilon_1$ and $\varepsilon_2$ simultaneously using a sum of Drude-Lorentz oscillators,\cite{dressel02a}
\begin{eqnarray}
\epsilon(\omega)=\epsilon_{\infty}+\sum_j\frac{\omega_{p,j}^2}{\omega_{0,j}^2-\omega^2-\imath\gamma_j\omega}
 \label{Eq:DrudeLorentz}
\end{eqnarray}
where $\omega_{0,j}$, $\omega_{p,j}$, and $\gamma_j$ are the peak frequency, the plasma frequency,
and the damping of the $j$th oscillator, and $\epsilon_{\infty}$ denotes the dielectric constant
at ``infinite'' frequency (i.e., above the measured region). The plasma frequency is related to the
spectral weight of a single Lorentzian,\cite{dressel02a}
$\int_{0}^{\infty} \sigma_1(\omega) d\omega = \omega_p^2/8$.

\begin{figure*}[tb]
     \includegraphics[width=1.7\columnwidth,clip]{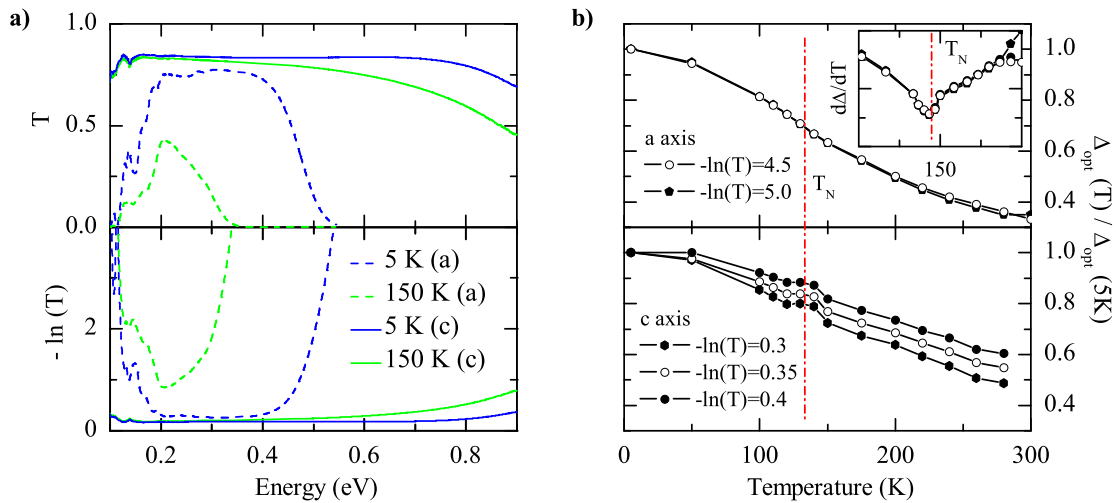}
    \caption{(Color online) (a) Transmittance $T$ (top panel) of a thin
    LaSrMnO$_{4}$ sample (d$\sim$70 $\mu$m) and $-\ln({\rm T}) \propto \alpha$
    for 5 and 150 K.\@
    (b) Evolution of the onset of the optical gap as determined from $-\ln({\rm T})=$const.
    }
    \label{fig5}
\end{figure*}

Using seven oscillators, we obtain an excellent description of both $\varepsilon_1^a$ and
$\varepsilon_2^a$, which clearly demonstrates the Kramers-Kronig consistency [see Fig.\ \ref{fig4}(a)].
The peak frequency of one of these seven oscillators is outside the measured region (dashed line).
It corresponds to the strong feature observed at about 7 eV by Moritomo {\em et al.} at room
temperature.\cite{moritomo95a}
The parameters $\omega_0$ and $\gamma$ of this strong oscillator have been assumed to be
independent of $T$. Since ellipsometry determines both $\varepsilon_1$ and $\varepsilon_2$
independently, the contributions of higher-lying bands to the measured region can be
fixed quite accurately. For the lowest three oscillators, the effective carrier concentration
$N_{\rm eff}$ obtained from the fit is displayed in Fig.\ \ref{fig4}(b).
With increasing $T$, the SW of peak (3) at 3.5 eV decreases by $\sim$20\%, while the SW of
peaks (1) and (2) at 1-2 eV increases by a factor of 2.
Comparing the $T$ dependence of $N_{\rm eff}$ below
about 3 eV as obtained either from the Drude-Lorentz fit [Fig.\ \ref{fig4}(b)] or from the direct integration of
$\sigma_1(\omega)$ (bottom panel of Fig.\ \ref{fig3}), the former is even stronger because it separates
contributions from higher-lying bands. The precise determination of $N_{\rm eff}$ is important for the
comparison with the kinetic energy (see Sec.\ V C below).

We find the same trend in the $c$ direction. The change of the SW of $\sim$10$\%$ obtained from direct integration
of $\sigma_1^c(\omega)$ may be influenced by a broadening or shift of the peak (since it is close to the
edge of the measured frequency range) or by a change in the background originating from higher-lying bands.
In a fit based on Eq.\ (\ref{Eq:DrudeLorentz}), we found a larger change in SW of $\sim$20$\%$.

\subsection{Transmittance and gap}

Figure \ref{fig5}(a) shows the transmittance $T$ and $-\ln(T) \propto \alpha(\omega)$ from 0.1 to 0.9 eV
for both $a$ and $c$ polarizations. Here, $\alpha(\omega)$ denotes the absorption coefficient.
The calculation of $\sigma_1(\omega)$ additionally requires the knowledge of the reflectivity $R$ in this
frequency range.
The transmittance is a very sensitive probe in order to determine the onset $\Delta_{\rm opt}$ of the optical gap.
For this purpose, $\alpha$ and $\sigma_1$ can be used equivalently.
Assuming a reasonable value of $R=0.15-0.2$, we find $\sigma_1^a \sim 1 (\Omega$cm$)^{-1}$
at 0.5 eV at 5 K, more than 2 orders of magnitude smaller than at 2 eV.\@ Thus, the data in
Fig.\ \ref{fig5} only show the very onset of excitations across the gap.
The absorption below about 0.2 eV can be attributed to weak (multi)phonon excitations.

From linear extrapolation of $-\ln(T)$ to zero, we find $\Delta^a_{\rm opt}=0.40-0.45$ eV
and $\Delta^c_{\rm opt}>0.9$ eV at 5 K.\@ In order to monitor the temperature dependence of the
onset of the gap, we solve the equations $-\ln(T^a)=m_a$ and $-\ln(T^c)=m_c$ with $m_a$=5.0 and 4.5
and $m_c$=0.40, 0.35, and 0.30, respectively, where $a$ and $c$ denote the polarization direction.
The results are shown in Fig.\ \ref{fig5}(b). At 300 K, we find $\Delta^a_{\rm opt}$ = 0.1-0.2 eV.\@
The redshift of the onset of the gap with increasing $T$ can be attributed to either a shift of the
peak frequency or an increase of the bandwidth.
According to the fit results, peak (1) shifts by about 0.3 eV from 5 to 300 K [see Fig.\ \ref{fig4}(a)].
This shift may originate from thermal expansion and electron-phonon coupling,\cite{cardona05aSSC,cardona05bRMP}
giving rise to a change of the effective crystal field.
The bandwidth $W$ may change due to either a change of the lattice parameters or of the spin-spin correlations
($W$ is reduced in the antiferromagnetically ordered state).
The plot of the derivative $d\Delta^a_{\rm opt}/dT$ in the inset of Fig.\ \ref{fig5}(b) and the lower panel clearly show that
$\Delta_{\rm opt}(T)$ changes its slope at $T_N$=133 K (independent of the choice of $m_a$ and $m_c$),
which reflects the behavior of the lattice constants.\cite{senff05a}

\section{Multiplet calculation}

One may expect that a local multiplet calculation for a single 180$^\circ$ Mn-O-Mn bond yields a reasonable
assignment of the CT ($d^4p^6 \! \rightarrow \! d^5p^5$) and MH excitations ($d^4d^4\! \rightarrow \! d^3d^5$)
of the undoped Mott insulator LaSrMnO$_4$.
For simplicity, we neglect hybridization; thus, the two-site states are a simple product of two single-site
states. This affects the excitation energies and the matrix elements.
Thus, we obtain renormalized parameters, e.g., an effective value of $U_{\rm eff}$, which has to be kept in mind
for comparison with results from other techniques. The selection rules are not affected by hybridization,
as stated in the Introduction.
We calculated $\sigma_1^a(\omega)$ and $\sigma_1^c(\omega)$ by evaluating the matrix elements
between all multiplet states.
For more details, we refer to the Appendix and Ref.\ \onlinecite{goesslingthesis}.

\begin{figure*}[tb]
     \includegraphics[width=1.8\columnwidth,clip]{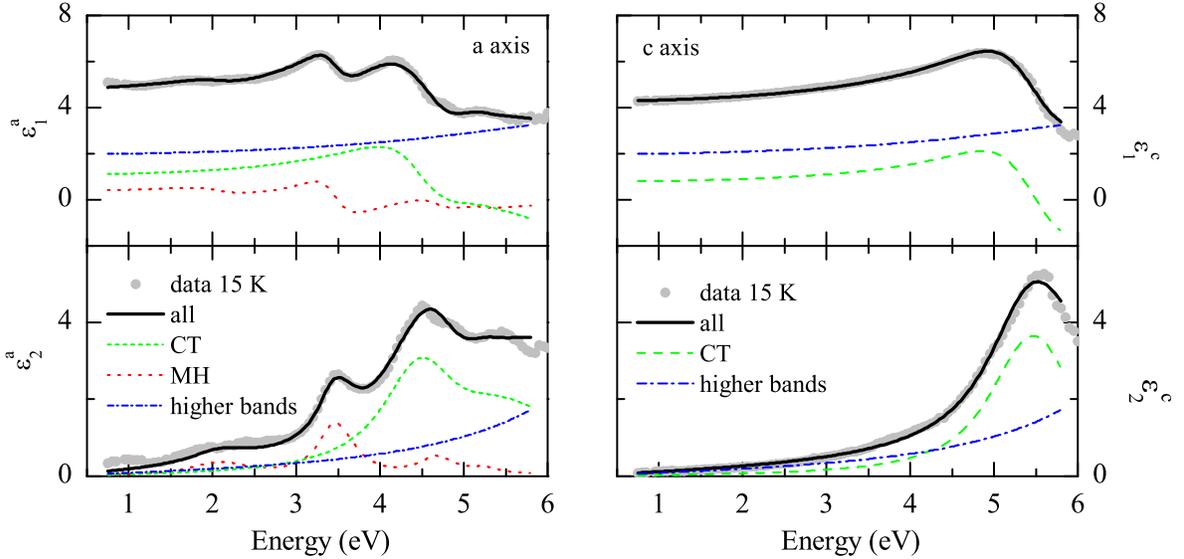}
    \caption{(Color online) Comparison between $\epsilon$ as obtained from the
    multiplet calculation and the measured data at 15 K.
    }
    \label{fig6}
\end{figure*}

The multiplet calculation takes into account the Coulomb interaction and the crystal-field splitting.
The former is described by the Slater integrals $F^0$, $F^2$, and $F^4$.\cite{sugano70a,haverkortPHD}
We use only two parameters, $F^0$ and the reduction factor $r$ in $F^k(d^n)\!=\!r F^k_{HF}(d^n)$ for $k$=2 and 4,
where $F^k_{HF}(d^n)$ denotes Hartree-Fock results for a free $d^n$ ion.
For the $d^n$ states, the tetragonal crystal field is parametrized by $10\,Dq$, $\Delta_{t2g}$, and $\Delta_{eg}$,
representing the splitting between $t_{2g}$ and $e_{g}$ levels and the splitting within these levels, respectively.
The CT energy is given by $\Delta_l \! = \! E_0(d^5) +  E(p^5) - E_0(d^4) -  E(p^6)$ for $l=a$ or $c$,
where $E_0(d^n)$ is the lowest energy of the $d^n$ multiplets. The $p^5$ states are assumed to be degenerate.
These seven electronic parameters (see Tab.\ \ref{tab:multiplets1}) determine all peak frequencies and the relative weight
of different peaks within one polarization. A constrained fit of the experimental data requires seven more parameters
(for the width, the absolute value, and higher-lying bands, see the Appendix).
Altogether, we use 14 parameters,
i.e., much less than those in the Drude-Lorentz model (22 parameters only for the $a$ direction, see above).
Moreover, the discussion of the peak assignment below will show that the pronounced structure of
the experimental spectra provides severe constraints for the electronic parameters.
The spectra with the lowest $\chi^2$ are plotted in Fig.\ \ref{fig6}. The overall agreement is excellent.
We find $F^0$=1.2 eV and $r$=0.64, resulting in $U_{\rm eff}$=2.2 eV and Hund's coupling $J_H$=0.6 eV (cf.\ Ref. \onlinecite{UJH}).
The energy levels of the $d^3$, $d^4$, and $d^5$ multiplets are shown in Fig.\ \ref{fig7}
as a function of the crystal-field parameters.

\begin{table}[tb]
\caption{\label{tab:multiplets1}Effective electronic parameters obtained from the multiplet
calculation by fitting the optical data for $T$=15 K.\@
The factor $r$ is dimensionless; all other units are in eV.   }
\begin{ruledtabular}
\begin{tabular}{lcccccr}
 $F^0$ & $r$ & $10\,Dq(d^4)$ & $\Delta_{t2g}(d^4)$ & $\Delta_{eg}(d^4)$ & $\Delta_a$ &
 $\Delta_c$\\[0.1em]
 \hline\\
  1.20 & 0.64 & 1.20 & 0.2 & 1.4 & 4.51 & 4.13\\
  \end{tabular}
  \end{ruledtabular}
\end{table}

\section{Discussion}

Our first goal is to disentangle CT excitations $d^4p^6\rightarrow d^5p^5$ and
MH excitations $d^4d^4\rightarrow d^3d^5$.
As discussed in the Introduction, the spectral weight of MH excitations is weaker.
In layered LaSrMnO$_4$, the interlayer Mn-Mn hopping is strongly suppressed and MH excitations
can be neglected in the $c$ direction.
This is the main reason for the pronounced anisotropy observed experimentally and suggests
the following interpretation: the peak at $\sim$5.6 eV in $\varepsilon_2^c(\omega)$ is a CT excitation
and the same holds true for the strong excitations in the same energy range in $\varepsilon_2^a(\omega)$.
The weaker features below $\sim$4 eV in $\varepsilon_2^a(\omega)$ are MH excitations.
The detailed analysis discussed below will support this assignment.

\subsection{Charge-transfer excitations}

\begin{figure*}[tb]
    \includegraphics[width=1.80\columnwidth,clip]{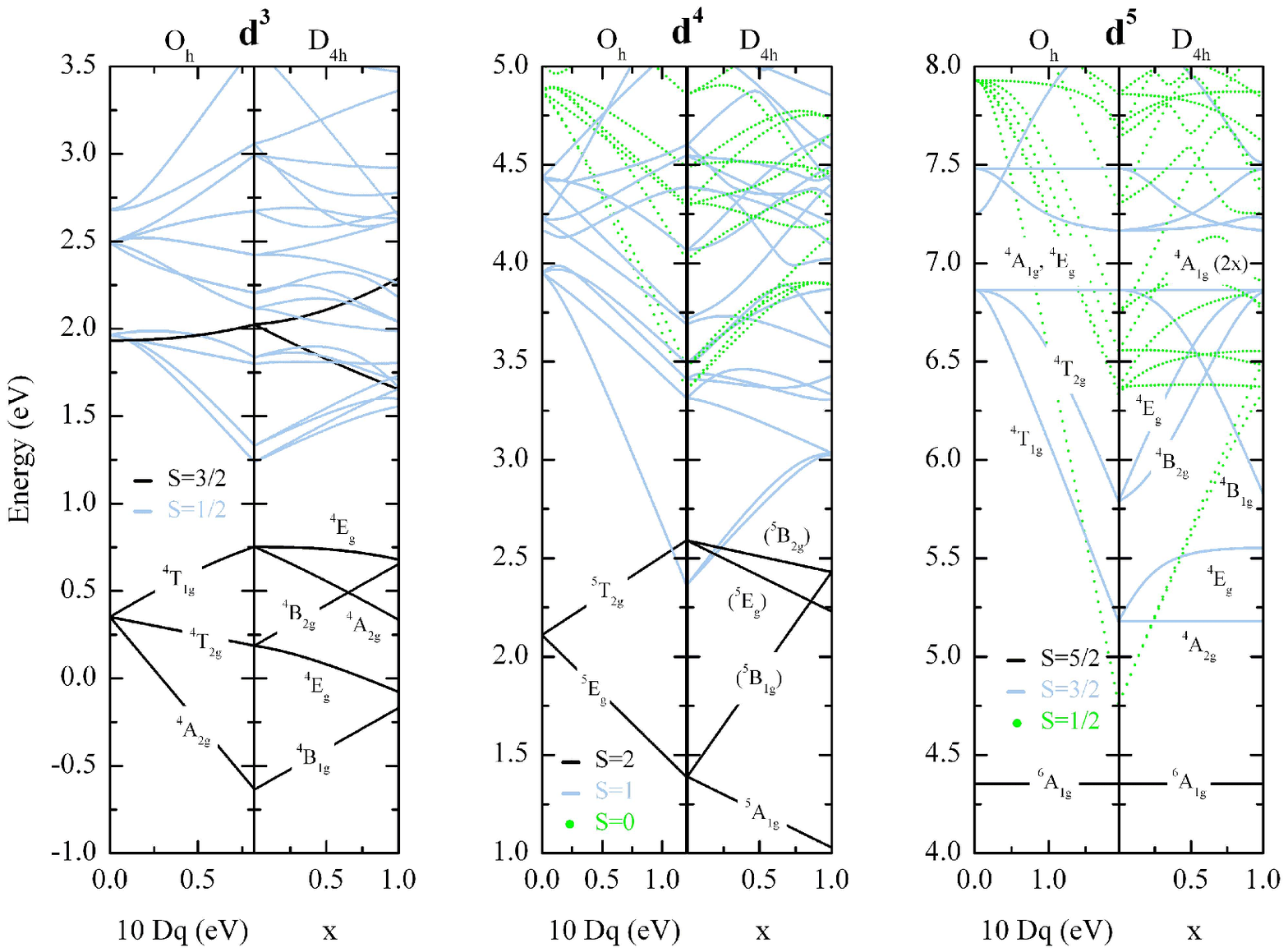}
    \caption{(Color online) Energy level diagrams from the multiplet calculation for the $d^3$, $d^4$, and $d^5$
    configurations as a function of $10\,Dq$ in $O_h$ and as a function of $x$ in $D_{4h}$ for a fixed value of $10\,Dq$.
    The control parameter $x$ denotes the strength of $\Delta_{eg}$ and $\Delta_{t2g}$, $x$=1 represents the full strength
    used for the spectra shown in Fig.\ \ref{fig6}. The low-lying multiplets are labeled by their irreducible representations
    in $D_{4h}$, those being not relevant for the optical transitions are shown in brackets. The electronic fit parameters
    are summarized in Table \ref{tab:multiplets1} (for more details, see the Appendix and Ref.\ \onlinecite{goesslingthesis}).
    }
    \label{fig7}
\end{figure*}

In the $c$ direction, we attribute the whole spectral weight to a CT peak at
5.6 eV and to the onset of higher-lying processes.
To get an idea about the initial and final states of this CT excitation, we start from
the strong crystal-field limit (i.e., neglect electron-electron interactions).
In the $d^4$ ground state, the tetragonally distorted MnO$_6$ octahedra
indicate that three electrons occupy the $xy$, $yz$, and $zx$ orbitals and the fourth electron occupies the
$d_{3z^2-r^2}$ orbital,\cite{wu04a,senff05a} whereas $d_{x^2-y^2}$ remains unoccupied
(see Fig.\ \ref{fig:lasr_multiplet_comicFM}). This is supported by our multiplet calculation.
In the initial state $d^4p^6$, all O $2p$ orbitals are occupied; thus, the excitation and its
selection rules are dominated by the $d^5$ part of the final state $d^5p^5$.
According to the spin selection rule, only $d^5$ states with $S$=5/2 or 3/2 can be reached
from the $d^4$ $S$=2 ground state. Following Hund's rule, the lowest $d^5$ state corresponds
to a high-spin $S=5/2$ multiplet, in which the five $3d$ orbitals are equally occupied.
For the orbital selection rules, one thus has to consider the overlap between the O $2p$ orbitals
and the initially unoccupied $d_{x^2-y^2}$ orbital.
This is only finite along the $a$ direction but zero along $c$. Therefore, we cannot identify
the peak at 5.6 eV in $\varepsilon_2^c$ with the {\it lowest} CT transition.
The second lowest CT excitation in the strong crystal-field limit corresponds to a transfer of one
electron from O $2p$ into the degenerate $d_{xz}/d_{yz}$ orbitals, i.e., to a final state
with $3d^5$ $S$=3/2. The overlap is finite, both in the $a$ and $c$ directions.
This assignment of the peak at 5.6 eV in $\varepsilon_2^c$ is supported by our multiplet
calculation (cf.\ Fig.\ \ref{fig6}).
However, the calculation resolves the contributions of different multiplets to
the peak at 5.6 eV, and it gives the relative weight of the different CT bands.

The calculated spectrum (see Fig.\ \ref{fig6}) shows only one strong CT band at 5.6 eV
in $\varepsilon_2^c$, while in $\varepsilon_2^a$ another strong CT band is observed at 4.5 eV.\@
The latter results from an excitation into the lowest $d^5$ final state with $S$=5/2 $(^6A_{1g})$.
The next $d^5$ states ($^4A_{2g}$, $^4E_{g}$, and $^4B_{1g}$) are found 0.9 - 1.5 eV above the
$^6A_{1g}$ multiplet (see Fig.\ \ref{fig7}).
The calculation shows that the second lowest excitation actually corresponds roughly to a transfer
from O $2p$ to $3d_{xy}$ [see the sketch of $d^5(^4A_{2g})$ in Fig.\ \ref{fig:lasr_multiplet_comicAFM}(b)],
but excitations to $d^5(^6A_{1g})$ and $d^5(^4A_{2g})$ are forbidden along $c$ by the orbital selection rule,
as discussed above in the strong crystal-field limit for the lowest CT absorption.
Only the transitions to $d^5(^4E_{g})$ and $d^5(^4B_{1g})$ are allowed in the $c$ direction
[see sketch of the $d^5$ states in Figs.\ \ref{fig:lasr_multiplet_comicAFM}(a) and (c)].
These constitute the peak at 5.6 eV.\@

Along $a$, the peak frequency is somewhat lower (5.5 eV), reflecting the excitation to the $^4A_{2g}$ final state
and a small anisotropy of the CT energy. From the fit of the entire spectrum, we find
$\Delta_a=4.51$ eV and $\Delta_c=4.13$ eV.\@ This is reasonable because the in-plane O(1) site and the apical
O(2) site are crystallographically different.\cite{senff05a,merz06b}

We emphasize that the assignment is unique.
The lowest CT excitation (hopping from O $2p$ to $3d_{x^2-y^2}$) is the strongest one
in $\sigma_1^a(\omega)$=$(\omega/4\pi)\varepsilon_2^a$.
Interpreting the peak at 3.5 eV
or even the 2 eV band as the lowest CT excitation does not yield sufficient weight around 4.5 eV in $\sigma_1^a(\omega)$.
Moreover, the selection rules show unambiguously that the peak at 5.6 eV along $c$ corresponds to excitations to
the $d^5(^4E_{g})$ and $d^5(^4B_{1g})$ multiplets, which are 1.2 - 1.5 eV above the lowest $d^5$ state.
Thus, we conclude that the lowest CT energy is $\gtrsim$ 4 eV.\@

The increase of spectral weight in $\sigma_1^c(\omega)$ with increasing temperature (see top panel of Fig.\ \ref{fig3})
can partially be attributed to the decrease of $\sim$0.5\% of the Mn-O(2) distance $d_c$ from 20 to
300 K.\cite{senff05a} With $t_{pd}\propto d_c^{-4}$ (Ref. \onlinecite{harrison99a}) and
$\sigma_1^c \propto d_c^2\,t_{pd}^2 \propto d_c^{-6}$, the decrease of $d_c$ can only account for a change in SW
of $\sim$3\%, in contrast to the observed gain of $\sim$10\%, see Fig.\ \ref{fig3}.
This may reflect a change in the occupation of the $3z^2$-$r^2$ orbitals, see Sec.\ V C.

\subsection{Mott-Hubbard excitations}

The observed value of $\sigma_1^a$ of a few 100 $(\Omega$cm$)^{-1}$ around 2 eV is typical for a Mott-Hubbard
absorption band in transition-metal oxides,
e.g., in RTiO$_3$ or RVO$_3$.\cite{arima93a,okimoto95a,imada98a,tsvetkov04a}
The SW around 2 eV cannot be attributed to local $dd$ excitations (crystal-field excitations), which
are parity forbidden within a dipole approximation. A finite SW is obtained by the simultaneous excitation of a
symmetry-breaking phonon, typically resulting in $\sigma_1$ of
only a few $(\Omega$cm$)^{-1}$.\cite{rueckamp05a,ballhausen62a}

First, we discuss the spin selection rule. The ground state is a $d^4$($^5$A$_{1g}$)$d^4$($^5$A$_{1g}$) state
(see Fig.\ \ref{fig7}), i.e., $S$=2 on both sites. If the spins are parallel ($S_i^z$=2 on both sites $i$=1, 2),
only fully spin-polarized states with $S(d^3)=3/2$ and $S(d^5)=5/2$ can be reached by the transfer of one electron
with $S$=1/2. For antiparallel spins ($|S_1^z|$=2, $S_2^z$=-$S_1^z$), we can reach final states with
$S(d^3)$=3/2 and $S(d^5)$=3/2 or 5/2 ($|S^z|$=3/2 in both cases).

\begin{figure}[bt]
\begin{center}
    \includegraphics[width=0.9\columnwidth,clip]{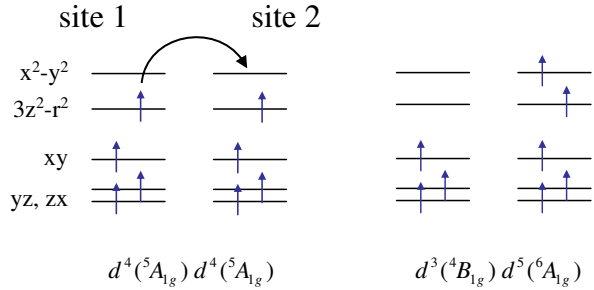}
    \caption{Sketch of the lowest Mott-Hubbard excitation in LaSrMnO$_4$ in the
    strong crystal-field limit, i.e., configuration mixing is neglected.
    In the final state, the $d^5$ site is in a high-spin $S$=5/2 configuration.
    This excitation is assigned to the feature observed around 2 eV in $\sigma_1^a(\omega)$.
    The spectral weight is largest for parallel alignment of neighboring spins.
    }
    \label{fig:lasr_multiplet_comicFM}
    \end{center}
\end{figure}

\begin{figure*}[bt]
\begin{center}
    \includegraphics[width=1.4\columnwidth,clip]{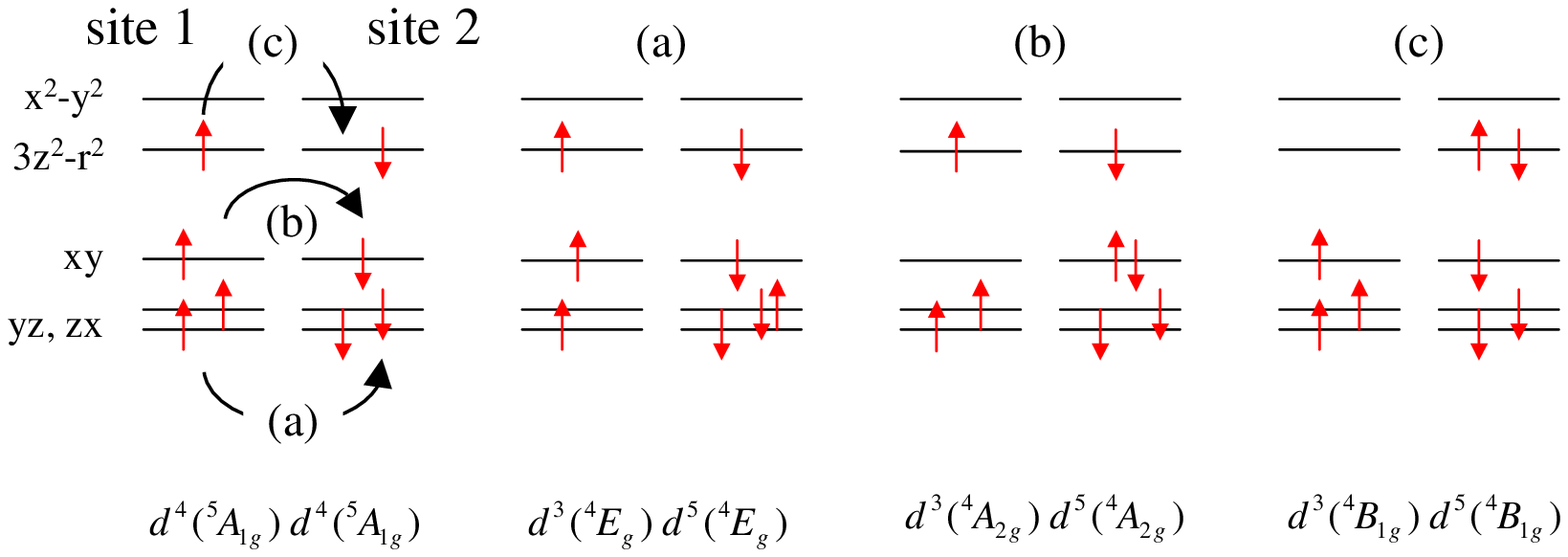}
    \caption{Sketch of Mott-Hubbard excitations to final states with $S(d^5)$=3/2.
    These excitations are degenerate in the strong crystal-field limit and
    are assigned to the peak at 3.5 eV in $\sigma_1^a(\omega)$.
    The spectral weight is largest for an antiferromagnetic arrangement of neighboring spins.
    }
    \label{fig:lasr_multiplet_comicAFM}
    \end{center}
\end{figure*}

We start the peak assignment again from the strong crystal-field limit. The highest occupied orbital $d_{3z^2-r^2}$ on one site
has overlap to both $d_{x^2-y^2}$ and $d_{3z^2-r^2}$ on the other site. In contrast, hopping of an electron from $xy$, $yz$, or $zx$
is only finite to the {\it same} type of orbital. In LaSrMnO$_4$, this selection rule is
strict because the O octahedra are neither tilted nor rotated.
Thus, one expects only three different Mott-Hubbard peaks in the strong crystal-field limit:
(i) the excitation from $3z^2$-$r^2$ to $x^2$-$y^2$ with a high-spin $S(d^5)$=5/2 in the final state
(see Fig.\ \ref{fig:lasr_multiplet_comicFM}),
this transition has the lowest energy according to Hund's rule (as long as $\Delta_{eg}$ is not too large);
(ii) the excitation from either $3z^2$-$r^2$ or $xy$ or from the degenerate $yz$, $zx$ orbitals to the {\em same} orbital
on the other site [$S(d^5)$=3/2, see Fig.\ \ref{fig:lasr_multiplet_comicAFM}],
these excitations have the same energy in the strong crystal-field limit because the orbital quantum number is preserved
in the transition, and all the final states show the same spin quantum numbers;
(iii) again an excitation from $3z^2$-$r^2$ to $x^2$-$y^2$, but this time with $S(d^5)$=3/2.


The Coulomb interaction gives rise to configuration mixing\cite{sugano70a} and
lifts the degeneracy of the excitations collected in (ii).  However, the overall features of the result of our multiplet
calculation are reproduced rather well by the crude approximation of the strong crystal-field limit discussed above.
As shown in Fig.\ \ref{fig7}, the lowest $d^3$, $d^4$, and $d^5$ states have $^4B_{1g}$, $^5A_{1g}$, and $^6A_{1g}$
symmetries, respectively. Thus, the lowest MH excitation is from the $d^4(^5A_{1g})d^4(^5A_{1g})$ ground state
to a $d^3(^4B_{1g})d^5(^6A_{1g})$ final state (Fig.\ \ref{fig:lasr_multiplet_comicFM}).
This statement is valid over a wide range of parameters. This transition
is assigned to the broad feature observed around 2 eV in $\sigma_1^a(\omega)$ (see Fig.\ \ref{fig6}).
The width is attributed to the large bandwidth of the $x^2-y^2$ band. In the Drude-Lorentz fit described in the
previous section, this excitation corresponds to peaks (1) and (2) [see Fig.\ \ref{fig4}(a)]. The fact that this feature is
not well described by a single Lorentzian can be attributed to band structure effects of the broad $x^2-y^2$ band.
Note that a similar ''fine structure`` was observed for the very similar lowest optical excitation in
LaMnO$_3$.\cite{kovaleva04a}
Due to the high-spin character of the $d^5$ final state, the spectral weight is largest for parallel alignment of
neighboring spins (see sketch in Fig.\ \ref{fig:lasr_multiplet_comicFM}). As discussed above, a $d^5$ final state
with $S$=5/2 and $S^z$=3/2 is also possible for antiparallel spins in the antiferromagnetic state below $T_N$=133 K,
but one expects a reduced SW.\@
Figure \ref{fig2} clearly shows an increase of the SW around 2 eV with increasing $T$,
in agreement with our assignment.

Both the $d^3$ and the $d^5$ configurations show several multiplets that are less than $\sim$1.5 eV above the lowest states.
For the identification of the next higher-lying excitations, we thus have to consider the orbital selection rule.
For the $d^4$ ground state, we find $\Gamma_{d^4}=A_{1g}$ (see Fig.\ \ref{fig7}), where $\Gamma$ denotes an irreducible
representation of the point group $D_{4h}$.
The matrix elements $\langle d^5_i | a^{\dagger}_{\tau} | d^4_{k}\rangle \langle d^3_j| a_{\tau'}|d^4_{k'}\rangle$
are only finite for $\Gamma_{d^5}\otimes\Gamma_{a^{\dagger}_{\tau}}\supset A_{1g}$ and
$\Gamma_{d^3}\otimes\Gamma_{a_{\tau'}}\supset A_{1g}$.
For excitations with $\tau=\tau'$, i.e., hopping within the same type of orbital (see sketch in Fig.\ \ref{fig:lasr_multiplet_comicAFM}),
we find $\Gamma_{d^5}=\Gamma_{d^3}$. According to Fig.\ \ref{fig7} we can thus attribute the peak at 3.5 eV in $\sigma_1^a(\omega)$
to excitations with the final states $d^3(^4E_{g})d^5(^4E_{g})$, $d^3(^4A_{2g})d^5(^4A_{2g})$,
and $d^3(^4B_{1g})d^5(^4B_{1g})$ [the higher-lying $d^3(^4E_{g})d^5(^4E_{g})$ transition has negligible weight],
which roughly correspond to the hopping of an electron within the $d_{xz,yz}$, $d_{xy}$, and $d_{3z^2-r^2}$ orbitals, respectively.
These excitations are degenerate within the strong crystal-field limit. According to the multiplet calculation, the splitting
is only small, giving rise to one pronounced feature at 3.5 eV.\@ Compared to the feature observed around 2 eV, the SW
at low temperatures is larger at 3.5 eV because three different processes contribute and because of the spin selection
rule.
Since the final states have $S(d^5)$=3/2, the SW of these transitions is largest for the antiparallel alignment  of
neighboring spins. According to the Drude-Lorentz fit of the previous section, the feature at 3.5 eV indeed loses weight
with increasing temperature [peak (3) in Fig.\ \ref{fig4}(b)]. The loss of $\sim$20\% from 15 to 330 K is not as strong as
the gain of the lowest transition. The direct integration of $\sigma_1^a(\omega)$ from 2.8 to 4 eV even yields a slight
gain of $\sim$3\% (see bottom panel of Fig.\ \ref{fig3}). As discussed above, this difference can be attributed to a change
of the background which can be resolved by the Drude-Lorentz fit. However, a precise quantitative determination of the
change of SW appears to be difficult in this frequency range, e.g., the background contribution of the CT
transitions may have been underestimated in the Drude-Lorentz fit.
The temperature dependence is discussed in more detail in the next section.

In comparison with the processes contributing to the peak at 3.5 eV, the excitation from $3z^2$-$r^2$ to $x^2$-$y^2$
with $S(d^5)$=3/2 is lower in Coulomb energy, but it costs about $\Delta_{eg}$. This can be identified with
the SW above $\sim$4 eV in the MH contribution (see Fig.\ \ref{fig6}).
Since $\tau \neq \tau'$, the orbital selection rule allows also for transitions to final states with
$\Gamma_{d^3} \neq \Gamma_{d^5}$; thus, different multiplets contribute. However, these are
difficult to separate from the CT excitations observed in the same energy range.

Our assignment is very well compatible with a number of different results.
In LDA+$U$ calculations,\cite{park01a,wu07a} the highest occupied band has mainly a Mn $3d_{3z^2-r^2}$ character (hybridized
with O $2p$ bands) and the lowest unoccupied band is a Mn $3d_{x^2-y^2}$ band.
Also, our interpretation of the peak at 3.5 eV is in agreement with the LDA+$U$ results.\cite{park01a,wu07a}
The x-ray data of Kuepper {\it et al.}\cite{kuepper06b} also suggest that the highest occupied states mainly have Mn character.
Our results support the very similar interpretation of the lowest optical excitation in LaMnO$_3$ in terms of a Mott-Hubbard
peak.\cite{jung97a,quijada,kovaleva04a,grenier05a}

\subsection{Temperature dependence and kinetic energy of the low-energy high-spin transition}

The superexchange interaction between spins on neighboring Mn sites arises from the virtual hopping of electrons
between the two sites. The intersite excitations probed in optical spectroscopy are the real-state counterpart of these virtual excitations.
Therefore, the superexchange constant $J$ is related to the spectral weight of the optical excitations.

In total, superexchange in LaSrMnO$_4$ favors antiparallel spins, but there is a ferromagnetic contribution
$J_{\rm FM}$, which corresponds to the lowest optical excitation to a high-spin $d^3(^4B_{1g})d^5(^6A_{1g})$
final state\cite{feiner99a} [peaks (1) and (2) of the Drude-Lorentz fit, see Fig.\ \ref{fig4}(a)].
The relation between the spectral weight or $N_{\rm eff}$ of this excitation and $J_{\rm FM}$ has been
derived\cite{oles05a,khaliullin05a} for the $d^4$ compound LaMnO$_3$.
The $c$ direction of LaMnO$_3$ with ferro-orbital order of $3x^2$-$r^2$ is equivalent to the $a$
direction of LaSrMnO$_4$ with $3z^2$-$r^2$ orbitals.
Adopting the formalism\cite{kovaleva04a} for LaSrMnO$_4$, the effective carrier concentration $N_{\rm eff}$,
the in-plane kinetic energy $K$, and the ferromagnetic contribution to superexchange $J_{\rm FM}$ are related by
\begin{eqnarray}
N_{\rm eff}&=&\frac{m (2d_a)^2}{\hbar^2}\,K \\
K &=&\frac{3}{8} J_{\rm FM} \langle\vec{S}_i\vec{S}_j+6\rangle\\
J_{\rm FM}&=&\frac{t^2}{5 E}
\end{eqnarray}
where $2d_a$=3.786 \AA\ is the Mn-Mn distance,\cite{senff05a}, $m$ the free electron mass,
$i$ and $j$ denote nearest-neighbor Mn sites,
$t=(pd\sigma^a)^2/\Delta_a$ the effective Mn-Mn hopping amplitude,
and $E= U_{\rm eff}-3J_H+\Delta_{eg}$ represents the excitation energy of the virtual intermediate state.
We determined $E$ from the weighted peak frequencies of peaks (1) and (2),
$E=(N_{\rm eff}^{(1)}\,E^{(1)}+N_{\rm eff}^{(2)}\,E^{(2)})/N_{\rm eff}$ = 2.10 eV at 15 K and
1.76 eV at 330 K.\@ The nearest-neighbor spin-spin correlation is given by
$\langle\vec{S}_i\vec{S}_j\rangle \! \rightarrow \! -4$ for $T \! \ll \! T_N$ and by
$\langle\vec{S}_i\vec{S}_j\rangle\! \rightarrow \! 0$ for $T \! \gg \! T_N$.
Using $\Delta_a$=4.5 eV from our analysis, we find $t \! = \! (pd\sigma^a)^2/\Delta_a \approx $0.6 eV.\@
Using $t$=0.55-0.65 eV, we derive $K$(15 K)=$\frac{3}{4}J_{\rm FM}$(15 K) = 0.021-0.030 eV
and $K$(330 K)=$\frac{9}{4}J_{\rm FM}$(330 K) = 0.077-0.108 eV and finally
$N_{\rm eff}^{\rm AF}$(15 K)=0.040-0.056 and $N_{\rm eff}^{\rm para}$(330 K)=0.146-0.203
[see Fig.\ \ref{fig4}(b)].
Both at low and at high temperatures, the calculated values agree amazingly well with the
experimental results. This strongly corroborates our interpretation of the feature around 2 eV
with the lowest Mott-Hubbard excitation. Moreover, it suggests that the redistribution of weight
with temperature can be attributed mainly to a change of the nearest-neighbor spin-spin correlation
function. As discussed above for the $c$ direction, the change of the lattice constant may only
account for a change of $N_{\rm eff}$ of a few percent.

A more detailed analysis requires the knowledge of the temperature dependence of
$\langle\vec{S}_i\vec{S}_j\rangle$. In the three-dimensional antiferromagnet LaMnO$_3$, the change
of $N_{\rm eff}$ right at $T_N$ is more pronounced.\cite{kovaleva04a} The more gradual changes
observed in LaSrMnO$_4$ qualitatively agree with the expectations for a quasi two-dimensional compound,
in which $\langle\vec{S}_i\vec{S}_j\rangle$ is still significant above $T_N$.\cite{fleury70a}

\subsection{Crystal-field excitations and thermal expansion}

Up to 600 K, the $c$-axis lattice parameter shrinks with increasing temperature and the elongation of the
octahedra is reduced. As a measure for the deviation from undistorted octahedra, we consider the difference
between the apical Mn-O(2) and in-plane Mn-O(1) bond lengths, which amounts to $d_c - d_a \approx $ 0.39 \AA{}
at 20 K and $\approx$ 0.37 \AA{} at 300 K, i.e., it
changes by more than 5\%.\cite{senff05a} The anisotropy of the thermal expansion has been
interpreted as an indication for a change of the orbital occupation.\cite{senff05a}
The change of $d_a$ across $T_N$ may be rationalized in terms of the gain of magnetic energy
with increasing $J$ (induced by an increase in $t$), but the change of $d_c$ suggests an orbital
effect.
On the basis of near-edge x-ray-absorption fine structure data, Merz {\em et al.} claimed a 15\% occupation of $x^2$-$y^2$ orbitals
at room temperature\cite{merz06a} as well as a transfer from $x^2$-$y^2$ to $3z^2$-$r^2$ with
decreasing temperature.\cite{merz06b}

Daghofer {\it et al.}\cite{daghofer06a,daghofer06b} studied the competition of various exchange contributions.
For $\Delta_{eg}$=0, they find that $x^2$-$y^2$ is occupied on each site for the case of antiferromagnetic order.
The fact that $3z^2$-$r^2$ is favored instead is due to $\Delta_{eg}>0$. Using $\Delta_{eg} \approx 0.1$ eV
$\sim$ 1160 K, Daghofer {\it et al.}\cite{daghofer06b} find a significant redistribution of electrons from
$3z^2$-$r^2$ to $x^2$-$y^2$ with increasing temperature.
However, $e_g$ electrons in general are strongly coupled to the lattice; thus, one expects much larger
values, e.g., $\Delta_{eg}>$1 eV in LaMnO$_3$.\cite{rueckamp05a}
The transmittance is a very sensitive probe for low-lying crystal-field excitations,\cite{rueckamp05a}
but in LaSrMnO$_4$, we do not find any evidence for crystal-field excitations below the optical gap
(see Fig.\ \ref{fig5}). Moreover, our multiplet calculation yields $\Delta_{eg}$=1.4 eV (see Table
\ref{tab:multiplets1}), in agreement with the strongly elongated O octahedra.
More recently, Rosciszewski and Oles\cite{rosciszewski} pointed out that the $3z^2$-$r^2$ orbitals
are occupied for $\Delta_{eg}>0.1$ eV.

\begin{figure}[t]
\begin{center}
    \includegraphics[width=.95\columnwidth,clip]{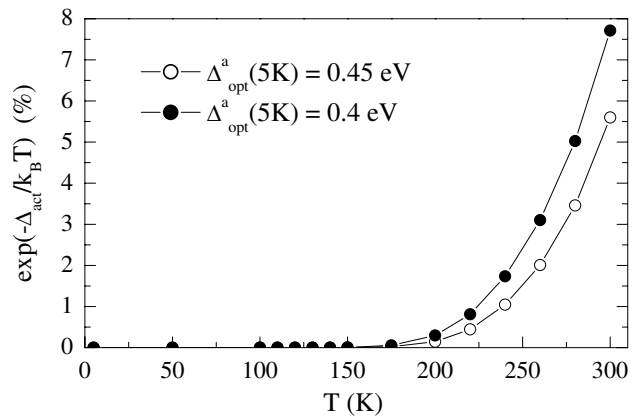}
    \caption{Estimate of the thermally activated population of the UHB using $\Delta_{\rm act}=(1/2)\Delta^a_{\rm opt}$
    and the temperature dependence of $\Delta^a_{\rm opt}$ depicted in Fig.\ \ref{fig5}(b).
    }
    \label{fig:boltzmann}
    \end{center}
\end{figure}

According to the Franck-Condon principle, optical excitations are very fast and probe the {\em static}
crystal-field splitting, i.e., for a frozen lattice. For the thermal activation, one has to consider
the minimal crystal-field splitting $\Delta_{eg}^{\rm rel}$ after relaxation of the lattice.
In optics, the peak frequency of a crystal-field absorption band corresponds to the static
value $\Delta_{eg}$ (neglecting the phonon shift of 50-80 meV required for breaking the parity
selection rule), whereas the onset of the absorption band can be identified with $\Delta_{eg}^{\rm rel}$.
Our transmittance data for $T$=5\,K clearly show that $\Delta_{eg}^{\rm rel}>$0.4 eV $\sim$ 4600 K.\@
Thus, we conclude that the thermal population of excited crystal-field levels, i.e., a {\em local} transfer
of electrons from $3z^2$-$r^2$ to $x^2$-$y^2$, is not sufficient to explain the anomalous thermal expansion.

The lowest electronic excitation corresponds to the optical gap $\Delta^a_{\rm opt} \approx$ 0.4 eV.\@
The optical absorption process corresponds to the
creation of {\em two} particles: an electron and a hole in a conventional semiconductor
or an ``empty'' site (here $d^3$) and a ``double occupancy'' (here $d^5$) in a Mott-Hubbard
insulator. With the Fermi energy lying in the middle of the gap, the thermal activation energy
for a {\em single} particle is only $\Delta_{\rm act} \! = \! (1/2) \Delta^a_{\rm opt} \approx$ 0.2 eV.\@
In contrast, a crystal-field excitation corresponds to a {\em bound} electron-hole pair,
i.e., an exciton, for which $\Delta_{\rm act} \! = \! \Delta_{eg}$.
As discussed above, the lowest Mott-Hubbard excitation with a $d^3(^4B_{1g})d^5(^6A_{1g})$ final state
corresponds to the transfer of an electron from a $3z^2$-$r^2$ orbital to the $x^2$-$y^2$ band,
similar to the local crystal-field excitations in the scenario proposed by Daghofer {\em et al.}\cite{daghofer06a,daghofer06b}
However, the gain in kinetic energy of the {\em delocalized} $d^5$ final state is essential to obtain
a small activation energy.
A rough estimate of the thermally activated population of the $x^2$-$y^2$ UHB is obtained from $\exp(-\Delta_{\rm act}/k_B T)$,
using $\Delta^a_{\rm opt}$=0.4-0.45 eV at 5 K, the temperature dependence of $\Delta^a_{\rm opt}$ depicted in
Fig.\ \ref{fig5}(b) and $\Delta_{\rm act}=(1/2)\Delta^a_{\rm opt}$. The result is shown in Fig.\ \ref{fig:boltzmann}.
With a population of more than 5\% at 300 K, the thermal activation into the UHB is a reasonable candidate for the
explanation of the anomalous thermal expansion, at least above 150 K.\@
Note, however, that the shrinkage of the $c$ axis is already pronounced at lower temperatures.

Irrespective of the mechanism, we may ask to which extent our data support such a transfer.
A change of the orbital occupation will slightly affect the orbital selection rule.
One may speculate that the 10\% {\em increase} of the SW of the CT excitations observed in the $c$ direction
(see top panel of Fig.\ \ref{fig3}) and the simultaneous 10\% {\em decrease} of SW in $\sigma_1^a(\omega)$ above
4 eV partially are due to a reduced occupation of $3z^2$-$r^2$ and a slightly enhanced occupation of $x^2$-$y^2$,
respectively. This may explain the very similar $T$ dependence of the $c$-axis lattice constant\cite{senff05a}
and $N_{\rm eff}^a$ above 4 eV.\@
We emphasize that such a change of the orbital occupation of a few percent cannot explain the much more
dramatic change of the SW of the lowest MH excitation around 2 eV.

Merz {\em et al.} reported a finite occupation of $x^2$-$y^2$ at 80 K.\cite{merz06b} This cannot be attributed
to a thermal population but may reflect the important role played by hybridization.
Another possibility for deviations from the regular orbital occupation is the disorder on the
La/Sr sublattice.
Each Mn ion has eight La/Sr neighbors, but due to La/Sr disorder, one may have to consider sites with a
local surplus of Sr (or La) ions. In Sr$_2$MnO$_4$, one finds $d^3$ Mn$^{4+}$ ions.
For a random La/Sr distribution, one expects that, e.g., about 1\% of the Mn ions have seven or eight Sr neighbors.
It is plausible that the $d^4$ configuration is not very stable at these sites,
and one may speculate whether in this case $x^2$-$y^2$ is preferable over $3z^2$-$r^2$ so that the
electron can spread out further to neighboring sites.
Typically, one expects that more loosely bound electrons will give rise to absorption features below
the optical gap.
Lee {\em et al.}\cite{lee07a} reported a peak at about 0.6 eV in LaSrMnO$_4$ at 10 K
based on a Kramers-Kronig analysis of reflectivity data. They attributed this peak to impurities or
O nonstoichiometry. Note that this peak at 0.6 eV has a maximum value of $\sigma_1(\omega)\! \approx \! 200$ $(\Omega$cm$)^{-1}$,
whereas the maximum of peak (1) at 1 eV is only about 60 $(\Omega$cm$)^{-1}$ at 15 K [see inset of Fig.\ \ref{fig2}(b)].
This difference may point at a certain sample dependence of the data, supporting an interpretation in terms
of impurities or disorder, but it may also reflect the problems of the Kramers-Kronig analysis.
The room-temperature data of Moritomo {\em et al.}\cite{moritomo95a} are in good agreement with our results.
Above, we attributed both peaks (1) and (2) to the lowest MH excitation of the regular $d^4$ compound.
The presence of two distinct peaks can be explained as a band structure effect. A very similar
splitting has been observed\cite{kovaleva04a} in LaMnO$_3$.
We emphasize that the spectral weight of peak (1) is only small, its interpretation does not affect
the assignment of all other peaks discussed above.

\section{Conclusion}
Using a local multiplet calculation, we analyze $\sigma_1^a(\omega)$ and $\sigma_1^c(\omega)$ of LaSrMnO$_4$
from 0.75 to 5.8 eV.\@ We arrive at a detailed peak assignment in terms of the multiplet splitting of
charge-transfer and Mott-Hubbard excitations. We obtain an excellent description of the pronounced anisotropy
and of the relative spectral weights of the different absorption bands.
Applying the selection rules of an {\rm effective} Mott-Hubbard model, the behavior of the lowest electronic
excitations is described very well.
In particular, the strong change of the spectral weight of the lowest excitation as a function of temperature
can be attributed to the change of the nearest-neighbor spin-spin correlations $\langle\vec{S}_i\vec{S}_j\rangle$
in the Mott-Hubbard case.
We thus conclude that LaSrMnO$_4$ can effectively be discussed as a Mott-Hubbard insulator.
In this sense, the lowest optical excitation is from a $d^4(^5A_{1g})d^4(^5A_{1g})$ ground state to
a $d^3(^4B_{1g})d^5(^6A_{1g})$ final state.

The clear advantage of the multiplet calculation over, e.g., a much simpler discussion of the strong crystal-field limit
is that it yields a description of the entire spectrum as well as a full set of effective parameters.
We find $U_{\rm eff}$=2.2 eV and $J_H$=0.6 eV for the Coulomb interaction, $\Delta_a$=4.5 eV and $\Delta_c$=4.1 eV for
the charge-transfer energy, and $10\,Dq$=1.2 eV, $\Delta_{t2g}$=0.2 eV, and $\Delta_{eg}$=1.4 eV for the
crystal-field splitting of the $d^4$ configuration. The rather small value of $U_{\rm eff}$ is the result of our
effective model, which does not consider hybridization explicitly. Neglecting the hybridization most probably
gives rise to the main shortcoming of our approach: it fails to describe quantitatively the anisotropy of the
spectral weight, while the relative spectral weight of different peaks within $\sigma_1^a(\omega)$ is well described.

In transmittance measurements, we find $\Delta^a_{\rm opt}$=0.4-0.45 eV at 15 K for the onset of excitations across
the optical gap and 0.1-0.2 eV at 300 K.\@ Due to this small value, the thermal population of the UHB may explain the
anomalous shrinkage of the $c$-axis lattice parameter, at least above 150 K.

\section*{Acknowledgments}
We acknowledge the support by the DFG via SFB 608 and thank
J.~Baier for the specific heat measurements.

\section*{Appendix: Optical-conductivity calculation}
In the considered frequency range (0.75 - 5.8 eV),
$\sigma_1(\omega)= \sigma^{MH}+\sigma^{CT}+\sigma^{\rm hi}$
shows MH and CT excitations as well as the
onset of higher-lying bands such as La($5d$), Sr($5s$), Mn($4s$), and Mn($4p$).\cite{jung97a}
The latter correspond to the oscillator above the measured frequency range in the free fit (see Sec.\ III).
The optical conductivity is calculated in arbitrary units by the Kubo formula.\cite{dagotto02a,cuoco99a,dressel02a,vanderMarel05a}
For layered LaSrMnO$_4$, we assume $\sigma_1^{c,MH}=0$
because two Mn sites in different layers are far apart and the interlayer hopping
is only small. The Mott-Hubbard contribution $\sigma_1^{a,MH}(\omega)$ reads
\begin{eqnarray} \label{eq:sig2}
\sigma^{a,MH}(\omega)&=&\frac{4d_{a}^2}{N}\sum_{i,j,k,k'}\frac{\mathcal{M}_{i,j,k,k'}^{a,MH}}{E^{MH}_{i,j}}\,\delta(\omega-E^{MH}_{ij})
\end{eqnarray}
\begin{eqnarray}
\mathcal{M}_{i,j,k,k'}^{a,MH}&=&|\langle d^5_i d^3_j| \sum_{\tau,
\tau'}t_{\tau,\tau'}^{a,MH}
a^{\dagger}_{\tau}a_{\tau'}|d^4_{k}d^4_{k'} \rangle|^2,\\[1em]
&=&|\sum_{\tau, \tau'}t_{\tau,\tau'}^{a,MH} \langle d^5_i |
a^{\dagger}_{\tau} | d^4_{k}\rangle \langle d^3_j|
a_{\tau'}|d^4_{k'}\rangle|^2\nonumber.
\end{eqnarray}
Here, $d_a$ denotes the in-plane Mn-O(1) bond length, $N$=2 the two different spin orientations
in the antiferromagnetic ground state
($d^4_\uparrow d^4_\downarrow$, $d^4_\downarrow d^4_\uparrow$),
and the indices $i$, $j$, $k$, and $k'$ label the many-particle eigenfunctions, i.e., $d^n_{i}$ refers to the
$i$th eigenfunction of the $d^n$ configuration with eigenenergy $E_{i}(d^n)$,
$E^{MH}_{i,j}=E_i(d^5)+E_j(d^3)-2E_0(d^4)$,
$a^{\dagger}_{\tau}$ ($a_{\tau}$) creates (annihilates) an electron in the orbital $\tau$,
where $\tau$ and $\tau'$ label the complex Mn $3d$ orbitals,
and the $dd$ hopping matrix $t_{\tau,\tau'}^{a,MH}$ indicates the overlap between two $d$ orbitals via a bridging
O ion. This matrix governs the orbital selection rules, it has been obtained from the Slater-Koster
tabular.\cite{slater54a}
The charge-transfer contribution $\sigma_1^{l,CT}(\omega)$ for $l$=$a$ or $c$ has been calculated
analogously.\cite{goesslingthesis}

For each excitation, the multiplet calculation gives the peak frequency $\omega_{0}$ and the spectral weight,
e.g.,
\begin{eqnarray} \label{eq:omegap}
\omega_p^2/8=\frac{4d_{a}^2}{N}\sum_{i,j,k,k'}\mathcal{M}_{i,j,k,k'}^{a,MH}/E^{MH}_{i,j} .
\end{eqnarray}
For the line shape, we assume a Lorentz oscillator [see Eq.\ (\ref{Eq:DrudeLorentz})].
The peak width $\gamma$ cannot be derived from our local model. As fit parameters, we employ $\gamma^{CT}$ for
all CT excitations and $\gamma^{MH}$ for the MH peaks
($\gamma = 2\gamma^{MH}$ for the peaks below 2.6 eV to account for the larger width of the
$x^2$-$y^2$ band). For $T$=15 K, we find $\gamma^{CT}$=1.10 eV and $\gamma^{MH}$=0.44 eV.\@
The onset of higher-lying bands is described by one oscillator with $\omega_0^{\rm hi}=8$ eV,
$\gamma^{\rm hi}=2.8$ eV, and $\omega_p^{\rm hi}=6.23$ eV at $T$=15 K, where only $\omega_p^{\rm hi}$
has been varied in the fit. The contribution to $\varepsilon_1$ arising from all excitations lying
at still higher energies is taken into account by $\varepsilon_\infty^{a}$=1.37 and
$\varepsilon_\infty^{c}$=1.51.
Since Eq.\ (\ref{eq:sig2}) is measured in arbitrary units, we use a global scaling factor $A$ for
$\sigma_1(\omega)$ (with $A$=24.83 from the fit).
The {\em anisotropy} of $\sigma_1$ is governed by the matrix elements. The hopping strength depends sensitively
on the distance, $pd\sigma \! \propto \! d^{-4}$ (Ref.~\onlinecite{harrison99a}).
For Mn-O distances\cite{senff05a} of $d_a$=1.88~\AA\ and $d_c$=2.28~\AA\, one expects
$pd\sigma^c/pd\sigma^a \approx 0.46$. Using this value, we underestimate the spectral weight
of the single peak observed at 5.6 eV in $\sigma_1^c(\omega)$.
Therefore, we scale $pd\sigma^c$ and $pd\pi^c$ with the common factor $A_c=1.96$.
This failure of describing the anisotropy of the spectral weight quantitatively most probably reflects
that we have neglected $pd$ hybridization 
(see Ref.~\onlinecite{goesslingthesis} for more details).

\end{document}